\documentclass[preprint,12pt]{aastex}

\begin{document}
\shorttitle{Gas Accretion at 13 Myr}
\shortauthors{Currie, T. et al.}
\title{Discovery of Gas Accretion Onto Stars in 13 Myr old h and $\chi$ Persei}
\author{Thayne Currie\altaffilmark{1}, Scott J. Kenyon\altaffilmark{1}, Zoltan Balog\altaffilmark{2,3},
 Ann Bragg\altaffilmark{4}, \& Susan Tokarz\altaffilmark{1}} 
\altaffiltext{1}{Harvard-Smithsonian Center for Astrophysics, 60 Garden St. Cambridge, MA 02140}
\altaffiltext{2}{Steward Observatory, University of Arizona,  933 N. Cherry Av. Tucson, AZ 85721}
\altaffiltext{3}{On leave from the Department of Optics and Quantum Electronics, University of Szeged, 
H-6720 Szeged, Hungary}
\altaffiltext{4}{Department of Physics, Bowling Green State University, Bowling Green, OH}
\email{tcurrie@cfa.harvard.edu}
\begin{abstract}
We report the discovery of accretion disks associated with 
$\sim$ 13 Myr-old intermediate/low-mass stars in h and $\chi$ Persei. 
Optical spectroscopy of $\sim$ 5000 stars in these clusters and a surrounding halo population reveal 
32 A-K stars with H$_{\alpha}$ emission. Matching these stars with 2MASS and optical 
photometry yields 25 stars with the highest probability of cluster membership and EW(H$_{\alpha}$) $\ge$ 5 \AA.
Sixteen of these sources have EW(H$_{\alpha}$)$\ge$ 10\AA.  
The population of accreting sources is strongly spectral type dependent: H$_{\alpha}$ emission 
characteristic of accretion, especially strong accretion (EW(H$_{\alpha}$) $\ge$ 10\AA),
 is much more prevalent around stars later than G0. 
Strong H$_{\alpha}$ emission from accretion is typically 
associated with  redder K$_{s}$-[8] colors. 
The existence of accreting pre-main 
sequence stars in h and $\chi$ Persei implies that circumstellar gas in some systems, especially those 
with primaries later than G5 spectral type, can last longer than 10-15 Myr.  
\end{abstract}
\keywords{Galaxy: Open Clusters and associations: Individual: NGC Number: 
NGC 869, Galaxy: Open Clusters and associations: Individual: NGC Number: NGC 884, 
accretion, accretion disks, planetary systems: protoplanetary disks}
\section{Introduction}
Young stars are born with massive, $\sim$ 0.01-0.1M$_{\star}$ (stellar mass), disks of gas and 
dust.  The disk viscously spreads, transporting angular momentum away from the star and mass 
onto the star.  Accretion onto the host star is identified from strong H$_{\alpha}$ emission; 
 typical mass accretion rates onto the star are $\sim$ 10$^{-8}$ M$_{\odot}$ yr$^{-1}$ 
for $\sim$ 1 Myr-old stars \citep{Hr98}.  After $\sim$ 5 Myr, fewer sources show strong H$_{\alpha}$ emission indicative 
of accretion, and accretion rates are typically much lower 
($\sim$ 10$^{-9}$ M$_{\odot}$ yr$^{-1}$) than at earlier ages.  
By $\sim$ 10 Myr, few sources show signs of active accretion \citep{Si05}.
 
The timescale for accretion to cease and for nebular gas in circumstellar disks to disperse 
has important implications for planet formation. Massive planets may form in gas-poor/free conditions 
(\citealt{Ch07, Kw03}).  However, gas giant planet formation requires that 
circumstellar gas lasts longer than the time for a $\sim$ 10-15 M$_{\oplus}$ protoplanetary 
core to capture a large gaseous envelope ($\sim$ 10$^{6}$-10$^{7}$ 
years; \citealt{Ik00, Pn05}).  Though the formation timescale is much shorter, 
the disk instability model (e.g. \citealt{Bo05}) for Jovian planet formation 
also requires a massive gas disk.  Circumstellar gas is also necessary
 for planetary migration (see \citealt{GT80, Wa97}), which may explain close-in giant planets ('hot Jupiters'), 
and circularization of the orbits of terrestrial planets.  
 A spread in the time for accretion to cease and for gas to disperse may 
then lead to diverse planetary systems.

While previous studies show that few stars accrete gas after $\sim$ 10 Myr \citep{Si05}, 
the bulk properties of the longest-lived accreting systems are not well constrained due to
small number statistics.  There are $\sim$ 5 known actively accreting stars with ages 
$\gtrsim$ 10 Myr, including PDS 66, AK Sco, and St 34 (Mamajek et al. 2002; Reipurth et al. 1996; \citealt{Wh05}). 
Well-known 10-20 Myr old clusters such as NGC 7160 (\citealt{Si05}) and Sco-Cen \citep{Ch05} have $\lesssim$ 
200 known members.   Because accretion at ages $\gtrsim$ 10 Myr
is rare ($\sim$ 2\%; \citealt{Si05}), a larger ($\gtrsim$ 1000-2000 sources)
 ensemble of 10-20 Myr sources spanning a wide range of spectral types is 
required to investigate the oldest accretion disks.

The double cluster, h and $\chi$ Persei --  13 $\pm$ 1 Myr old, d=2.34 kpc \citep{Bk05, Sl02, Ke01} --
provides an excellent opportunity to study the properties of a statistically significant sample of 
the longest-lived accretion disks.  With $\gtrsim$ 5000 members  
in the cluster and surrounding halo population \citep{Cu07a}, robust constraints on gas 
accretion are possible even if the fraction of accreting sources is $\sim$ 1-2\%.

In this Letter, we report the discovery of 25 sources in h and $\chi$ Persei with 
evidence for active gas accretion.  
In \S 2 we identify these accreting sources by measuring 
 H$_{\alpha}$ equivalent widths and use quantitative spectral types 
 to select actively accreting h and $\chi$ Per members from a sample of 
$\sim$ 5000 optical spectra.  Analysis of our sample in \S 3 reveals that 
the population of accretors is almost solely comprised of stars with spectral types later than 
G5.   We compare the strength of accretion signatures 
to dust emission in \S 4, finding a weak correlation between the H$_{\alpha}$ 
equivalent width and IR excess.  We conclude with a 
brief summary and discuss future observations to investigate 
circumstellar gas accretion in young stars in more detail.  
These results show that circumstellar gas in at least stars 
later than $\sim$ G0 lasts $\gtrsim$ 10-15 Myr.

\section{Hectospec Observations/Hydra Archival Data of h and $\chi$ Persei and Sample Identification}
We obtained Hectospec \citep{Fa05} spectra of 4536 sources with V $\sim$ 16-19, J$\sim$ 14.25-16.25, 
and J-H $\sim$ 0-1.5 in $\chi$ Persei and the halo population surrounding both clusters
on the 6.5m MMT telescope at Whipple Observatory during September and November 2006. 
Each source was observed in three 10-minute exposures
using the 270 mm$^{-1}$ grating.  This configuration yields spectra at
4000-9000 \AA\ with 3 \AA\ resolution.   The data were
processed using the standard Hectospec reduction pipeline \citep{Fa05} and typically had S/N $\gtrsim$ 
30-50 at 5000 \AA.

We acquired additional spectra of 710 sources near h and $\chi$ Per with
the Hydra multifiber spectrograph \citep{Ba93} on the WIYN 3.5 m telescope at the Kitt Peak National
Observatory.  Hydra spectra were obtained during two observing runs in November 2000 and October 2001
and include stars with V $\sim$ 14-17 and J $\sim$ 12-14.5 (S/N $\approx$ 10-30 typically).  
We used the 400 g mm$^{-1}$ setting blazed at 42$^{o}$, with a resolution of 7 \AA\
 and a coverage of 3600-6700 \AA.  The standard IRAF task \textit{dohydra} was
used to reduce the spectra.  

To select candidate emission line stars, we measured spectral indices (I($\lambda$); 
see \citealt{BK02})
with the IRAF \textit{sbands}\footnote{IRAF is distributed by the National Optical
Astronomy Observatory, which is operated by the Association of
Universities for Research in Astronomy, Inc. under contract to the
National Science Foundation.} routine.  Because classical T Tauri stars have 
strong H$_{\alpha}$ emission and often have strong Ca II emission, 
we derived indices for H$_{\alpha}$, H$_{\beta}$, H$_{\gamma}$, H$_{\delta}$, 
and several Ca II features.  For the H I features, bandpasses with 
widths of 30 \AA\ ( H$_{\alpha}$) and 20 \AA\ (H$_{\beta}$, H$_{\gamma}$, and H$_{\delta}$) 
yield good results.  For Ca II, we centered bandpasses at 3933 \AA\ (20 \AA\ width)
 and at 8498, 8542, and 8662 \AA\ (20 \AA) for the IR triplet lines.
To identify sources with H$_{\alpha}$ emission, we compared 
the spectral index of H$_{\alpha}$ to H$_{\beta}$, which was 
less likely to be in emission and typically had high signal-to-noise. 

\begin{figure}
\epsscale{0.9}
\plottwo{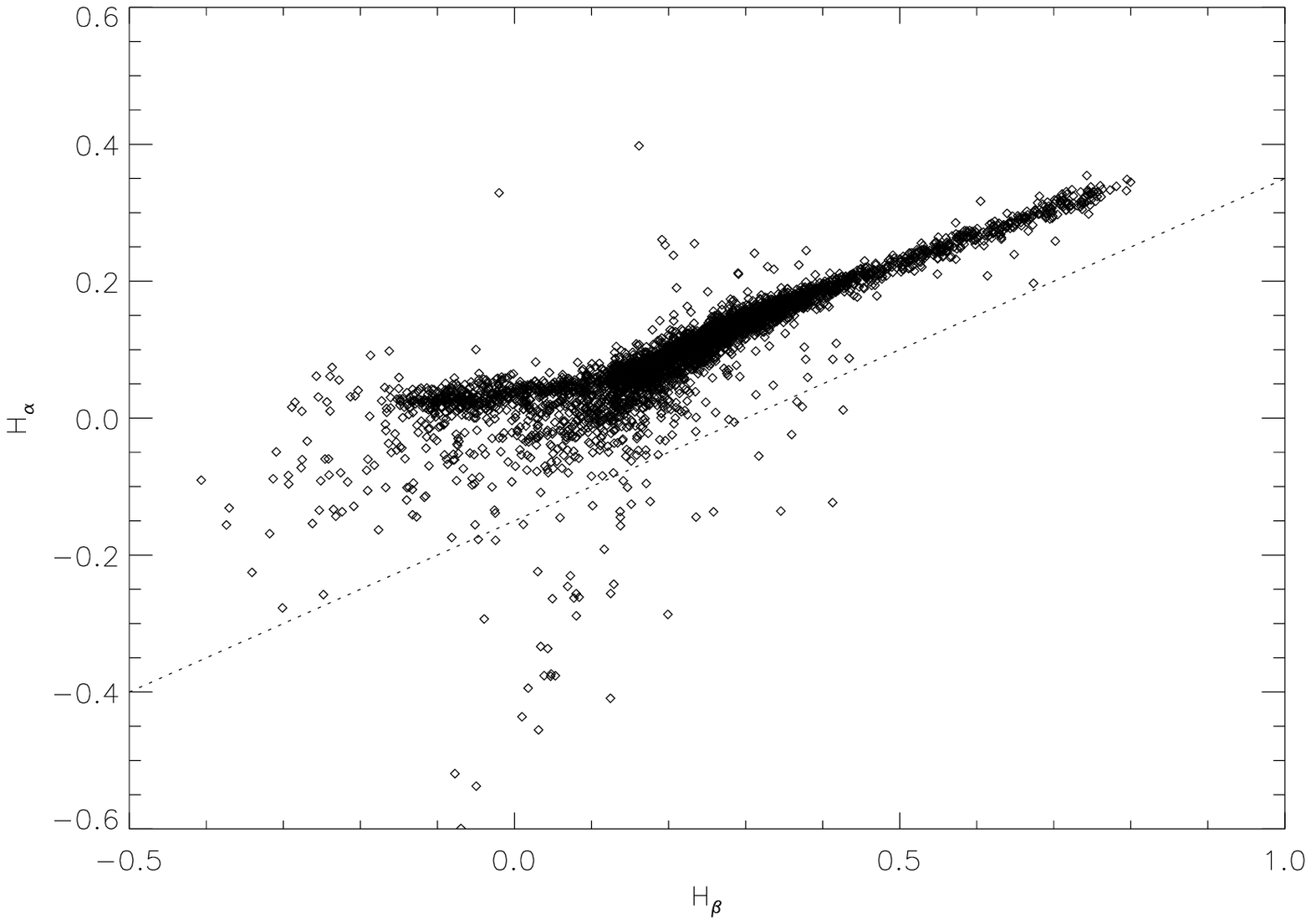}{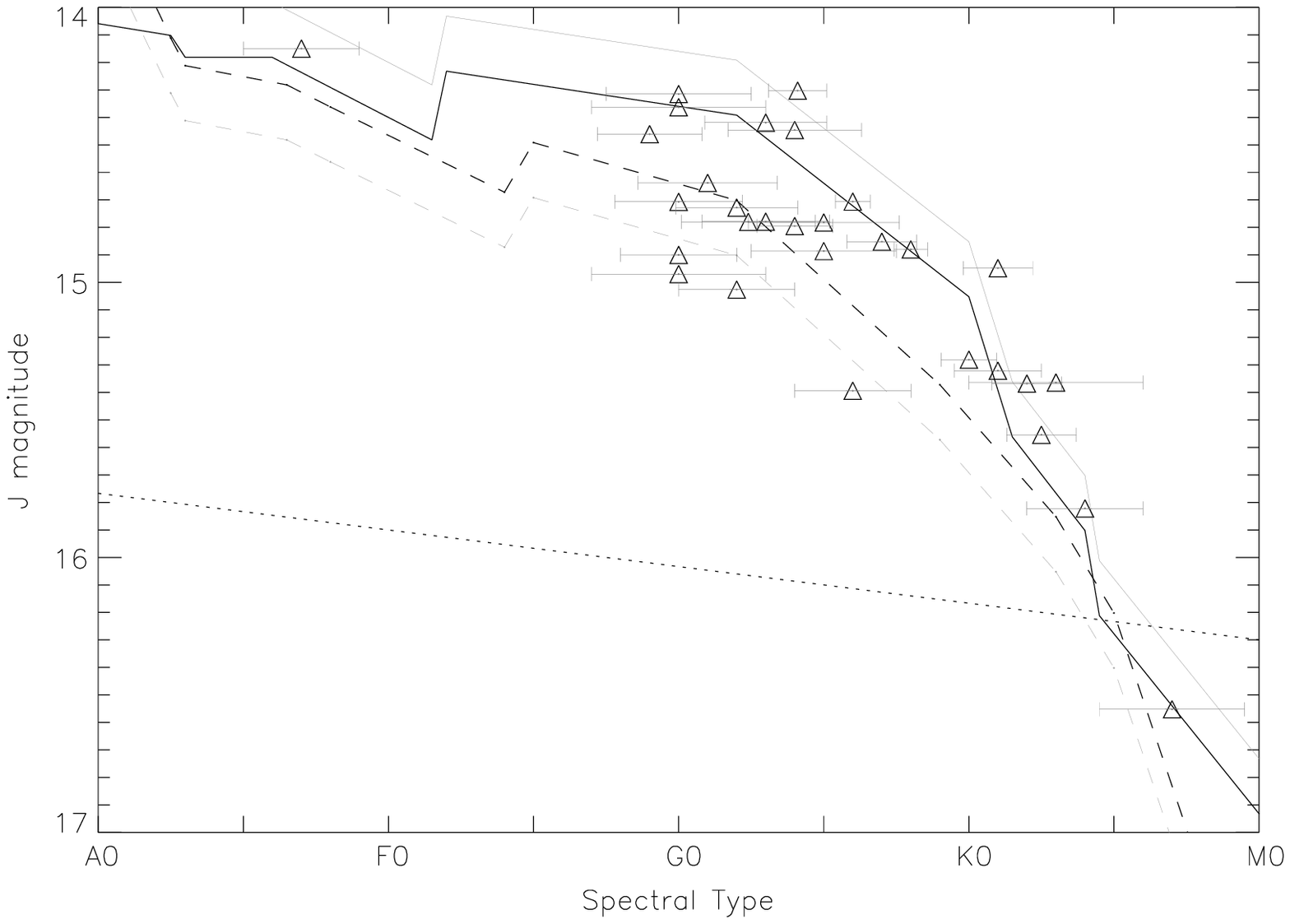}
\caption{(left) The H$_{\alpha}$ spectral index plotted against the H$_{\beta}$ index.  Sources lying below the dotted line 
were selected for emission signatures;  sources between the main distribution and the line typically had lower signal-to-noise.  
(right) Spectral type vs. J magnitude for h and $\chi$ Per members with H$_{\alpha}$ emission.
The dotted line indicates the faint limit of the 
 survey.  The 12 and 14 Myr isochrones (top and bottom thick lines)
are overplotted with 0.2 mag errors (top and bottom grey lines).  The errors in derived spectral type 
are overplotted as bars on the data points (triangles).}
\label{specv}
\end{figure}

The H$_{\alpha}$ and H$_{\beta}$ spectral indices for most stars
 lie in a narrow band on I(H$_{\beta}$) vs. I(H$_{\alpha}$) 
from $\sim$ (-0.2,0) to (0.7,0.4).  Accreting sources have 
anomolously small H$_{\alpha}$ index for a 
given H$_{\beta}$ index (Figure \ref{specv}a; see also \citealt{BK02} and \citealt{Br02}). 
Candidate accreting sources are those lying outside the main distribution of sources 
on I(H$_{\beta}$) vs. I(H$_{\alpha}$), below the line I(H$_{\alpha}$)=0.5$\times$ I(H$_{\beta}$)-0.15 
(Figure \ref{specv}a).  After removing background M giants, Be stars, and low signal-to-noise 
stars, 32 sources with clear H$_{\alpha}$ emission and 
moderate to high signal-to-noise spectra remain.

While each source clearly shows H$_{\alpha}$ emission, sources with 
weaker emission may not be accreting \citep{Wb03}.
Chromospheric activity can produce 
 very small H$_{\alpha}$ equivalent widths (EW (H$_{\alpha}$)$\lesssim$ 3-5 \AA).  
Accreting T Tauri stars typically EW(H$_{\alpha}$) $\ge$ 10-20 \AA, 
though low rates of accretion ($\dot{M}$ $\gtrsim$ 10$^{-8}$ M$_{\odot}$yr$^{-1}$) 
can produce EW(H$_{\alpha}$) $\sim$ 3-10 \AA\ for stars earlier than K5 \citep{Wb03, Bn05}.  
We consider sources with EW(H$_{\alpha}$) $\gtrsim$ 5\AA\ as potential accreting stars.
To distinguish more clearly between 
accreting sources and chromospherically active sources, we remeasured EW(H$_{\alpha}$)
 using the IRAF routine \textit{splot}, smoothing any low signal-to-noise spectra 
and fitting the line profile to a gaussian.  Typically, \textit{splot} derived only slighly smaller 
EW(H$_{\alpha}$) than \textit{sbands} ($\sim$ a few \AA). 
Sources with EW(H$_{\alpha}$) $\ge$ 10\AA\ are 'strong' accretors.  
Consistent with observations of T Tauri stars in Taurus-Auriga and other clouds 
(e.g. \citealt{Kh95}), no sources show evidence of CaII emission.

To derive quantitative spectral types for these 32 sources, 
we used spectral indices for $H_{\beta}$, $H_{\gamma}$,
 $H_{\delta}$, the G band (4305 \AA), and Mg I (5175 \AA).
  We fit piecewise linear 
relationships between each spectral index and spectral type from 
the \citet{Ja84} standards, derived spectral types for each index, and 
adopted the median as the spectral type.  
For the 3 sources with H$_{\beta}$ in emission, we derived a 
spectral type from the 4 other indices without signs of accretion.  
The resulting spectral type distribution ranged from A7 to M0 
with typical uncertainties of $\sim$ 1-2 subclasses.
We compared our results to those using the quantitative spectral 
typing method of \citet{He04} and found the agreement to be 
excellent.

To identify stars likely associated with the clusters, 
 we used optical photometry
(\citealt{Ke01}) and 2MASS/IRAC photometry from \citet{Cu07a}.  We exploited 
the well-constrained age and reddening of the double cluster and derived the 
expected 2MASS J magnitude of members at  
each spectral type.  Sources within 0.2 mags of a band defined by the 12 and 14 Myr 
stellar isochrones (from \citealt{Si00}) were identified as  
members (see \citealt{Cu07a}).  Figure \ref{specv} shows the 30 sources consistent 
with membership: the other two sources were a late K star and 
a M0 star that are too cool to fall within the isochrone.
Although these two sources may lie at the cluster distance and 
have slightly different ages (e.g. 10 or 15 Myr) or 
extinctions, the clusters appear to have a small age spread 
 \citep{Me93, Ke01, Sl02, Bk05}.  Thus, the 30 sources shown 
in Figure \ref{specv}b are the most likely cluster 
 members.  Table 1 lists their properties.

Young T Tauri stars show a strong correlation 
between EW(H$_{\alpha}$) and IR excess from disks \citep{Kh95}.  
Stars in $\sim$ 1 Myr-old Taurus-Auriga with L band (3.8$\mu m$) excess (K-L $\ge$ 0.4) 
typically have stronger H$_{\alpha}$ emission (EW(H$_{\alpha}$) $\gtrsim$ 
10\AA) than those lacking excess.  Excess sources 
 have a range of EW(H$_{\alpha}$), which implies a range of accretion 
rates.  To search for a similar trend in our sample, we 
use the K$_{s}$-[8] color derived from the 2MASS/IRAC 
data \citep{Cu07a} as an indicator of excess emission instead of K$_{s}$-[3.6].  Eighteen of
30 sources with H$_{\alpha}$ emission are detected at [8] with errors $\le$ 0.2 mags.
Typically, sources without 8$\mu m$ photometry are fainter in K$_{s}$ ($\sim$ 14.25-14.75) than those 
with 8$\mu m$ detections (K$_{s}$ $\sim$ 13.5-14). 
\section{Results: Evidence for Accretion at $\sim$ 13 Myr}
\begin{figure}
\epsscale{0.8}
\plotone{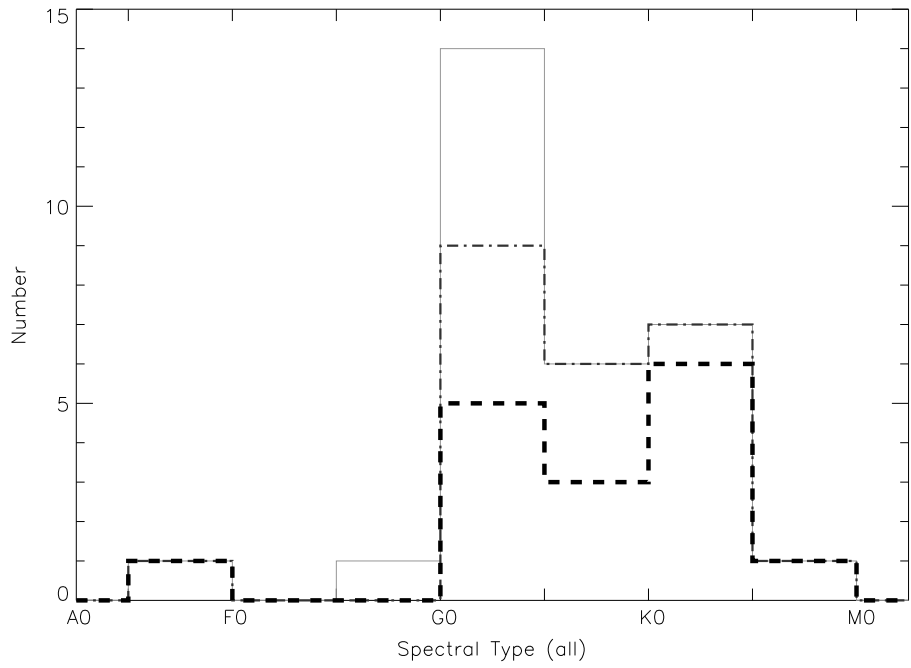}
\caption{Spectral type distribution of H$_{\alpha}$ emission sources (grey line), those with EW$\ge$5\AA\ (dash-dot line), and 
those with EW$\ge$10\AA\ (thick dotted line).  Sources with H$_{\alpha}$$\sim$5-10\AA\ are probably marginally accreting, 
those with H$_{\alpha}$ $\gtrsim$ 10\AA\ are consistent with the strength of H$_{\alpha}$ emission from 
Classical T Tauri stars.}
\label{specs}
\end{figure}

Most of the H$_{\alpha}$ emission stars in h and $\chi$ Per have G0 or later spectral types (Figure \ref{specs}).
Overall, 25/30 sources have EW$\gtrsim$5 \AA\ and thus 
are likely accreting circumstellar gas.  All fourteen sources later than G5 
 have EW $\gtrsim$5 \AA.  By contrast, 5/16 sources G5 and earlier have 
EW(H$_{\alpha}$) $\le$ 5\AA\ . 
The distribution of sources with EW(H$_{\alpha}$)  $\geq$ 10\AA\
is even more strongly biased towards later spectral types.  
Seven out of eight sources later than K0 have EW(H$_{\alpha}$) $\geq$ 10\AA\ 
while only 6/16 of the sources G5 and earlier have EW(H$_{\alpha}$) $\geq$ 10\AA. 
  Four of five sources with EW(H$_{\alpha}$)$\ge$ 20\AA\ are later than 
K0, and the sources with the strongest H$_{\alpha}$ emission (EW $\sim$ 47 \AA\ and 59 \AA) 
are a K7 and K4 star.  The atlas of both emission line and non-emission line stars will 
appear in the full spectroscopy survey 
paper (T. Currie et al. 2008, in progress).

The spectral type distribution of accreting sources 
 is comprised almost exclusively of sources later than G0.  
Because we do not have a complete sample of non-accreting sources at a given 
spectral type, we cannot accurately measure the frequency of accretors 
as a function of spectral type/mass.  Nevertheless, we can estimate the 
fraction of accreting sources in our sample by deriving how many of 
the 5145 (4536 from Hectospec, 609 from Hydra) sources with near-IR photometry
lie along the h and $\chi$ Per isochrone in J/J-H.  We follow \citet{Cu07a} and identify 
as cluster members 2040 sources within a band that extends 0.3 
mags fainter and 0.75 mags brighter (the binary locus) than the 
13 Myr isochrone, where the vast majority of non members lie below the 
isochrone.  We estimate the spectral types of sources 
from their J-H colors and divide the sample into two populations, 
earlier than F8 and later than F8,  and compute the frequency of 
accreting sources in each population.  

The frequency of accreting sources is low and is consistent with a spectral-type dependence.  
Overall, $\sim$ 1.2\% (25/2040) of sources consistent with cluster membership are likely accreting.
Through F8, 1/1042 (0.1\%) sources with spectra show evidence of accretion.  The 
frequency is 2.4\% (24/998) for later sources.
Spectral types for the non-accreting sources will provide better 
constraints on the frequency of long-lived accretion disks.
The low fraction of accreting stars in h and $\chi$ Per is broadly consistent with results from 
other studies.  \citet{Si05} and \citet{Si06} 
identified accretion in $\sim$ 40\% of stars in the 4 Myr-old Tr 37 Cepheus OB2 
subgroup and only one of 55 stars ($\sim$ 2\%) in the other Cep OB2 subgroup, 
NGC 7160 ($\sim$ 11.8 Myr old).  An accretion disk fraction of $\sim$ 1.2\% 
from our sample is consistent with a decline of accretion disk frequency with age.  

 To check these results, we consider possible selection biases.  If the
emission-line stars have significant optical veiling, their spectral
types appear earlier than fainter (J $<$ 16) non-accreting stars with
similar masses.  We then overestimate their relative frequency. However,
stars with large veiling have near-IR excesses, H--K$_s$ $\gtrsim$ 0.5
and K$_s$--[3.6] $\gtrsim$ 0.4-0.5 \citep{Kh95}. With H--K$_s$ $\lesssim$ 0.4-0.5
and K$_s$--[3.6] $\lesssim$ 0.3-0.4 \citep{Cu07a}, emission-line stars in
h \& $\chi$ Per have little or no near-IR excess and probably have negligible
veiling. If photometric errors move faint cluster stars outside the isochrone,
we underestimate the number of non-accreting stars. Including all stars
with J-H $\gtrsim$ 0.48 (F8; 2303 total) in the cluster sample yields a strict lower
limit to the accretion frequency, 1.1\% $\pm$ 0.2\%, which is still much
larger than the frequency for earlier type stars. Finally, we consider
the possibility that the H$\alpha$ emission flux associated with late-type
stars is undetectable around 1 mag brighter cluster stars with earlier
spectral types. With EW(H$\alpha$) $\sim$ 5--10 \AA\ for the late-type stars,
early-type stars with similar H$\alpha$ fluxes should have EW(H$\alpha$)
= 2--5 \AA, easily detectable on our spectra.  Thus, we conclude that
possible selection effects do not produce a false trend of increasing
accretion frequency among lower mass stars.

\begin{figure}
\epsscale{0.75}
\plotone{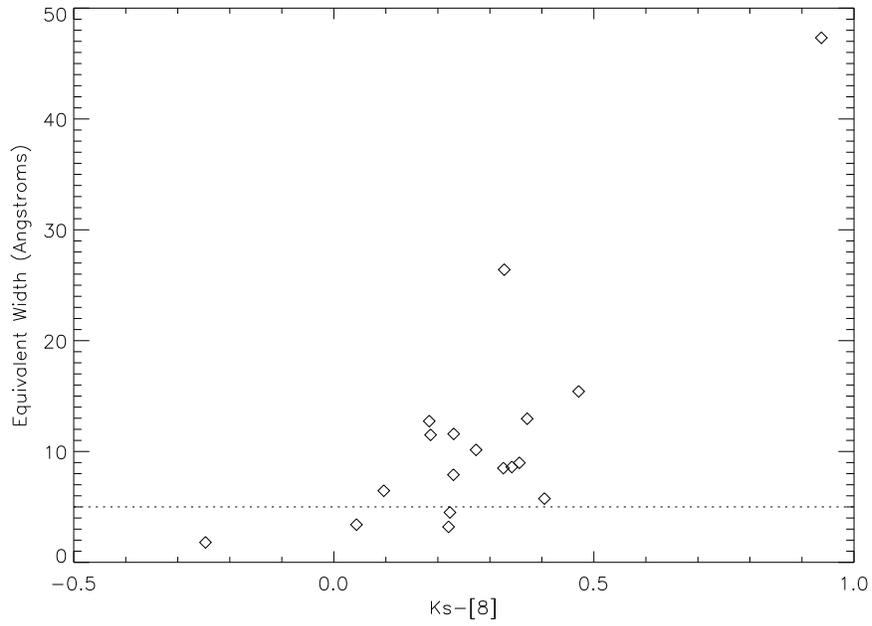}
\caption{H$_{\alpha}$ equivalent width vs. K$_{s}$-[8] 
color.  There is a weak correlation between the H$_{\alpha}$ width and IR excess.}
\label{eqwv}
\end{figure}

Our sample shows a $\sim$ 2-3$\sigma$ correlation between accretion and infrared excess. 
The majority of stars have negligible K$_{s}$-[8] excess (Figure \ref{eqwv}), but
some sources have K$_{s}$-[8]$\sim$ 0.4, the nominal cutoff for IR excess  
 in \citet{Cu07a} and \citet{Kh95}.  One source, the K7 star, has K$_{s}$-[8] $\sim$ 1 and
is clearly redder than a stellar photosphere.  Sources with 
larger H$_{\alpha}$ ($\ge$20\AA) have redder K$_{s}$-[8] 
colors than those with small/marginal H$_{\alpha}$ emission.  The Spearman rank correlation 
coefficient between K$_{s}$-[8] and H$_{\alpha}$ emission is r$_{s}$=0.60; this distribution 
has a low probability of being drawn from a random sample (p$_{d}$ = 0.9\%).  The distribution 
is less correlated for sources with chromospheric H$_{\alpha}$ emission (r$_{s}$=0.35, p$_{d}$ $\sim$ 22\%).  We thus 
find a $\sim$ 2-3 $\sigma$ correlation between IR excess and EW(H$_{\alpha}$)
for sources with H$_{\alpha}$ emission.  However, some debris disks in h and $\chi$ Per 
 \citep{Cu07b, Cu07c} have strong excesses at [8] and/or [24] without accretion.  
Therefore, while accretion may imply IR excess, IR excess need not imply accretion. 
\section{Summary \& Discussion}
We have discovered a population of stars with accretion disks in h and $\chi$ Persei. 
Of the 30 H$_{\alpha}$ emitters that are likely h and $\chi$ Per cluster/halo members, 25 
have EW(H$_{\alpha}$) consistent with circumstellar gas accretion.  The spectral types 
of accreting sources are almost all later than G0.  About $\sim$ 1.2\% of our sample 
shows evidence for accretion, consistent with the low fraction of accretors at 
$\sim$ 10 Myr found by \citet{Si05}.  We find that H$_{\alpha}$ emission sources with larger 
EW(H$_{\alpha}$) have redder K$_{s}$-[8] excess.  

While current survey points to a spectral type dependence on the accretion disk lifetime, 
 a full survey of h and $\chi$ Persei sources can provide stronger constraints on 
the properties of long-lived disks. 
Our sample is drawn heavily from $\chi$ Persei and the surrounding halo 
population.  Analyzing spectra of the slightly more massive/dense member of the 
Double Cluster, h Persei, will test whether the frequency of accreting stars 
is the same or depends on properties such as the mass and density of the cluster.
  A complete Hectospec survey of $\sim$ 
15,000 sources in the h and $\chi$ Per field is underway and will better characterize the 
population of accreting pre-main sequence stars.
\acknowledgements
We thank the anonymous referee for a timely and helpful report and 
Eric Mamajek for useful discussions regarding previously known stars with 
long-lived accretion disks.  TC and SK acknowledge support from the 
NASA Astrophysics Theory Program grant NAG5-13278 and TPF Grant NNG06GH25G.  
ZB received support from Hungarian OTKA Grants TS049872 and T049082.

\begin{deluxetable}{lllllllllllllll}
\tiny
\setlength{\tabcolsep}{0.01in}
\tablewidth{0pt}
\tabletypesize{\scriptsize}
\tablecolumns{10}
\tablecaption{H$_{\alpha}$ emission sources}
\tiny
\tablehead{{$\alpha$}&{$\delta$}&{ST}&{$\sigma$(ST, subclasses)}&{EqW(H$_{\alpha}$)}&{Accreting?}&{J}&{$K_{s}$}&{K$_{s}$-[3.6]}&{K$_{s}$-[8]}}
\startdata
2:22:21.62&57:04:00.5&G4&1.0&13.7&y&14.303&13.612&0.292&\ldots\\
2:22:39.53&57:15:42.9&G0&2.5&4.5&n&14.315&13.538&0.271&0.223\\
2:17:36.96&57:01:10.6&G0&3.0&1.8&n&14.363&13.697&\ldots& -0.247\\
2:17:55.15&56:56:03.1&G3&2.1&6.46&y&14.419&13.613&0.258&0.096\\
2:21:46.44&57:02:44.2&G4&2.3&5.75&y&14.446&13.697&0.349&0.405\\
\enddata
\tablecomments{Properties of sources with H$_{\alpha}$ emission (first five entries).  
Sources with EW(H$_{\alpha}$), $\ge$ 5\AA\ are identified as accreting sources. 
 All but the final entry are from Hectospec.  
Uncertainties in EW(H$_{\alpha}$) were $\sim$ 1-2 \AA\.  
All photometric measurements listed are 5$\sigma$ detections.  
Sources were typically brighter than the 10$\sigma$ limit in 
J (15.7) and K$_{s}$ (14.8) (see \citealt{Cu07a}).}
\end{deluxetable}

\end{document}